# QUALITY CONTROL, TESTING AND DEPLOYMENT RESULTS IN NIF ICCS

John P. Woodruff, Drew D. Casavant, Barry D. Cline, Michael R. Gorvad
LLNL, Livermore, CA 94550, USA

Abstract

The strategy used to develop the NIF Integrated Computer Control System (ICCS) calls for incremental cycles of construction and formal test to deliver a total of 1 million lines of code. Each incremental release takes four to six months to implement specific functionality and culminates when offline tests conducted in the ICCS Integration and Test Facility verify functional, performance, and interface requirements. Tests are then repeated on line to confirm integrated operation in dedicated laser laboratories or ultimately in the NIF. Test incidents along with other change requests are recorded and tracked to closure by the software change control board (SCCB). Annual independent audits advise management on software process improvements. Extensive experience has been gained by integrating controls in the prototype laser preamplifier laboratory. The control system installed in the preamplifier lab contains five of the ten planned supervisory subsystems and seven of sixteen planned front-end processors (FEPs). Beam alignment, timing, diagnosis and laser pulse amplification up to 20 joules was tested through an automated series of shots. Other laboratories have provided integrated testing of six additional FEPs. Process measurements including earned-value, product size, and defect densities provide software project controls and generate confidence that the control system will be successfully deployed.

## 1 INTRODUCTION

The Integrated Computer Control System (ICCS) software for NIF is being constructed by an iterative process that implements and tests specific functional increments as required by project management. Planned overlap between cycles of development allows management to level the effort of design and review, implementation, and test activities to utilize staff efficiently. Performance measures taken during each cycle show progress toward completion. Each increment is first tested in the ICCS testbed. Presently, this is followed by testing in the appropriate prototype laboratories.

## 2 ITERATIVE DEVELOPMENT CYCLES

Each planning cycle begins by identifying which requirements will be implemented for the increment. ICCS managers review project needs and select from the integrated project requirements documents those functions appropriate for the next development cycle. Availability of testbed resources, project risk resolution and prospects for integration with laboratory equipment influence the selection of requirements that are included.

The product of the requirements phase is an implementation plan: a document that defines the requirements, the changes that have been accepted into the software change request database, new features on the user interface, and dependencies on other subsystems. The implementation plan also establishes the schedule for the increment and defines which of the ICCS subsystems [1] will be deployed at the end of the increment. Earned value accounting allocates value for the product of each engineering phase, so the value accrues steadily through the construction of the increment.

Detailed design work on the increment's components begins when the implementation plan is complete. A review of software designs precedes implementation. Each of the subsystems in a deployment is reviewed prior to code implementation, with reviewers' action items tracked to closure. Included in the documentation presented in a review are Unified Modeling Language representations of the software classes to be built, interface definition language interfaces for CORBA-distributed objects, schemae for database tables, Buhr diagrams [2] showing concurrency, and user interface sketches.

Implementation and unit test are accomplished using an integrated development environment that incorporates version control and change management with the Ada compiler and the Java development suite.

When the several interacting programs that constitute a testable product release are completed, integration and deployment is performed by the configuration management team, independent of the implementers. The configuration manager performs quality control verification by witnessing compliance with the implementation plan. Requested software changes that were completed are documented via the change control database. The source code is stabilized by the version control tool. All versions are confirmed to be consistent, and the build process is repeated under configuration control. Completeness is confirmed by executing a set of integration tests that confirm interface consistency of all the communicating processes. The entire deployment is copied to the test environment and turned over to the test team.

## 3 TEST PLANNING AND EXECUTION

The test team's responsibility is to confirm that the delivered software correctly and robustly implements the requirements established for the increment. The testers design, execute, and document the results of extensive tests. Test plans and test procedures are written starting with the requirements, and accepted changes are documented in the implementation plan.

One or more tests are conducted for each deployment, with each test typically encompassing numerous test cases. For requirements traceability, test procedures identify the requirements being verified and the relevant procedural steps. Test procedures are redlined as required during test execution to provide an accurate basis for test expansion for subsequent deployments as well as regression testing.

Tests are conducted "offline" in a dedicated Integration and Test (I&T) facility, where representative laser hardware is available for extensive testing. In the I&T facility, the software can be exercised at the limits of its capabilities and error cases introduced to assure robustness. Testers also draft operations manuals and oversee preparation of configuration database instances that support activities in the I&T facility and in laser hardware prototyping laboratories.

Fully integrated deployments can be tested online in a laser laboratory. During both offline and online testing, incidents of noncompliance are recorded and analyzed for trends. Some tests expose defects in software: erroneous requirements, functional errors, errors of omission, and regressive failures are tracked. Software change requests are entered and managed by the SCCB. Hardware defects are documented and tracked using a similar process.

So far, over 650 test incidents have been documented; about three-quarters of these are software issues that result in software change requests (SCRs). Defects that prevent completion of critical testing or operations are classified as urgent and are repaired by the development team as quickly as possible. A patch to the deployment is then issued to fix the problem, and regression tests are performed. Software defect density to date has been approximately 2 functional defects per 1000 lines of code. The success rate for repaired defects is approximately 90%. New defects have been introduced in less than 10% of the patches.

## 4 OVERLAP OF SUCCESSIVE INCREMENTS

The foundation subsystems – the frameworks and the support layers where commercial off-the-shelf products are installed – lead application development by half a phase. The benefit of this tactic is that new framework functionality is specified ahead of need. Framework requirements are defined immediately after the application subsystems have completed their design reviews (in the preceding cycle), and the implementation of framework enhancements occurs while the succeeding application systems are being specified and designed. This tactic delivers (partially) tested frameworks to application developers just as they begin their intensive implementation activities.

Overlap between cycles is also exploited in the planning process. When subsystem design reviews are complete and the test plans and user documentation are available in draft form, the next cycle of preliminary planning starts. Thus, planners who learn of difficulties in realizing an implementation plan can adjust the next increment accordingly.

## 5 DEPLOYMENT VARIATIONS

The architecture of the ICCS is conducive to the phased implementation strategy because substantial amounts of functionality can be developed independently for the several functional subsystems. This allows distinct deployments of ICCS components as laboratories become available for online tests. Furthermore, all the subsystems rely on a common set of framework software [3] that itself is undergoing incremental refinement. Status propagation and display [4] are examples of functions that different subsystems implement independently while using the common ICCS framework.

These considerations lead to some tactical decisions about what steps of development should be executed for the different subsystems. Since the ICCS is a loosely coupled collection of subsystems, some deployments include only partial functionality: an example is the target diagnostic subsystem that is very loosely coupled to the laser controls. Standalone tests of target diagnostic functionality have been performed in the absence of other subsystems.

## 7 SOFTWARE CHANGE MANAGEMENT

Required functionality for an ICCS increment must respond to the project schedule. Management decisions based on the required functionality thus drive the development cycles. Another source of work for the development team is changes that are requested by a variety of stakeholders such as testers who report defects, developers who evolve internal interfaces, operators and project customers concerned with human factors. By convention, these two sources of work are kept distinct since increments of functionality are taken to be purely new code, while changes are expected to modify code that has already been tested. An SCCB manages the disposition of all requested changes.

SCRs are accepted into a managed database from any interested party; most arise either from defects exposed during testing or from evolution within the development team. The data associated with the SCRs allows the SCCB to know the status of each authorized change: what is the next step in satisfying the request, who is responsible for that activity, and when is completion expected? The final step in a change process is regression testing, which occurs when an incremental product is delivered. About 1100 SCRs have been considered since the SCCB was formed, and half of these are still incomplete.

## 6 EXPERIENCE WITH INCREMENTAL DEVELOPMENT

The ICCS has been under development since project inception in 1998; seven cycles of code release and test have been completed. The size of the successive releases has grown from 89 thousand source lines of code (KSLOCs) to a present inventory of 322 KSLOCs. The estimate of the product size at completion is about 1000 KSLOCs.

Early ICCS releases have been deployed to a variety of different destinations for testing. Because the NIF facility itself is not ready to receive control system software, the project has constructed two successive generations of I&T facilities. The present 2,400-square-foot facility houses 9 Unix workstations and 23 racks of electronics. Generally, one of each type of controls hardware module that will be used in the NIF is represented in the I&T facility. For example, rack-mounted equipment includes data and application servers, network switches, FEPs, programmable logic controllers, timing system components, motor controllers, emulators, and transient recorders. A lab table is used to mount motors, shutters, photodiodes, and other device points for functional testing.

Offline testing accomplished in the I&T facility is, for many subsystems, augmented by testing in dedicated laser subsystem prototyping laboratories. Controls deployed into these labs are integrated with laser hardware, shaking out interface issues and providing operations personnel the opportunity for controls validation and training. Integration tests in the prototyping labs are also used to proof installation and checkout procedures used for controls deployments into the NIF.

The most extensive online test to date has been accomplished in the Front-end Integration System Test (FEIST) laboratory, where the prototype laser preamplifier, input laser diagnostics sensor [5], and timing system are assembled. The control system installed in the FEIST lab contains five of the ten planned supervisory subsystems and seven of sixteen planned FEPs. Beam alignment, timing, diagnosis, and laser pulse amplification up to 20 joules was tested through an automated series of shots on the preamplifier. Other laboratories dedicated to wavefront control [6], pulsed power conditioning, and Pockels' cell testing have provided integrated testing of three additional supervisors and six additional FEPs.